\documentclass[12pt]{spieman}  

\usepackage{amsmath,amsfonts,amssymb}
\usepackage{graphicx}
\usepackage{setspace}
\usepackage{xurl} 
\usepackage{needspace}
\AddToHook{cmd/bibitem/before}{\Needspace{3\baselineskip}}
\newcommand{\ManuscriptSpacing}{1.0}

\graphicspath{{figures/}}

\newcommand{\Tc}{\bar{T}_2^*}
\newcommand{\etain}{\eta_{\mathrm{in}}}
\newcommand{\etaout}{\eta_{\mathrm{out}}}
\newcommand{\etaro}{\eta_{\mathrm{ro}}}
\newcommand{\ptilde}{\tilde p}
\newcommand{\Nph}{\mathcal{N}}

\title{Single-atom-based asynchronous photonic interconnect for scalable modular quantum computing}

\author[a]{J\'er\'emy Raskop}
\author[a]{Nadav Kandel}
\author[a]{Geva Arwas}
\author[a]{Yaniv Amichy}
\author[a]{Yaron Jarach}
\author[a]{Tal Kanonich}
\author[b]{Andrei Militaru}
\author[b]{Johannes Fink}
\author[a,c,*]{Barak Dayan}
\affil[a]{Quantum Source Labs Ltd, Ness Ziona, Israel}
\affil[b]{Institute of Science and Technology Austria (ISTA), Am Campus 1, 3400 Klosterneuburg, Austria}
\affil[c]{AMOS and Chemical and Biological Physics Department, Weizmann Institute of Science, 76100 Rehovot, Israel}

\begin{document}
\maketitle

\begin{abstract}
  Scaling quantum computation beyond the capacity of a single quantum processing unit requires quantum interconnects between modular processors. Optical photons are natural carriers for distributing entanglement between these processors. Most loss-resilient protocols use photonic Bell-state measurements based on the linear-optics type-II fusion gate. The resulting entanglement rate scales quadratically with each processor's typically low photon-delivery probability. Here we analyze a memory-assisted quantum interconnect using a near-deterministic, robust photon--atom controlled-$Z$ gate via a single atom trapped in a high-finesse cavity. Detecting and measuring a photon from one processor heralds entanglement between that processor and the atom. This entanglement is preserved while the process repeats with the second processor until the second photon is detected and measured. Reading out the atomic qubit finalizes the entanglement between the processors. As the entanglement is mediated by the atom, the photons from both processors do not need to be indistinguishable, removing a major source of infidelity. Furthermore, by removing the simultaneous photon-arrival requirement, the protocol allows the entanglement rate to scale linearly rather than quadratically with photon-arrival probability over a wide parameter range. We derive entanglement rates under realistic parameters, accounting for decay of the atom's entanglement with the first processor and for decoherence caused by unheralded photon interactions. The nanosecond-scale of the gate and read-out operations leads to orders-of-magnitude entanglement-rate gain over linear optics, removing a key bottleneck in modular quantum computing.
\end{abstract}

\keywords{quantum interconnect, Bell-state measurement, modular quantum computing, quantum memory, cavity quantum electrodynamics, photon--atom gate}

{\noindent \footnotesize\textbf{*}Corresponding author: Barak Dayan, \linkable{barak@qs-labs.com}}

\begin{spacing}{\ManuscriptSpacing}

  \section{Introduction}
  Scaling quantum computation beyond the capacity of a single quantum processing unit (QPU) will likely require modular architectures with quantum interconnects linking individual processors~\cite{monroe2014large,nickerson2014freely,barral,caleffi,sinclair2025,main}. By providing nonlocal couplings between QPUs, interconnects allow error-correcting codes to span multiple locally connected processors. This can increase the attainable code distance and support high-rate codes, expanding the fault-tolerant design space~\cite{strikis2023quantum,sutcliffe2025distributed}. For logical computation, the same connectivity enables transversal entangling gates between remote logical blocks~\cite{stack2026transversal} and simplifies lattice-surgery routing and fault-tolerant compilation~\cite{guinn2023codesigned,litinski2022active}. It also supports heterogeneous architectures with task-specialized QPUs~\cite{stein2025hetec,menssen2026,yoder2025tour}. The rate and fidelity of remote entanglement provided by the quantum interconnect are therefore important system-level resources alongside local-gate speeds and fidelities within each QPU~\cite{ramette2024fault,marqversen2026}.

\begin{figure}[tb]
    \centering
    \includegraphics[width=0.75\linewidth]{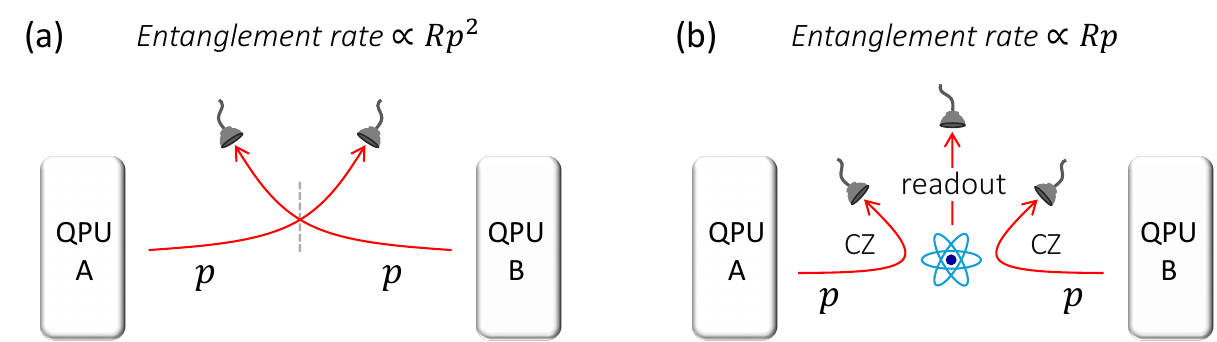}
    \caption{\textbf{Scaling of the entanglement rate.}
    (a) In conventional type-II fusion, quantum processing units A and B must each successfully deliver a photon in the same attempt. If each does so with probability $p$ at attempt rate $R$, the entanglement rate scales as $Rp^2$. (b) Using a photon--atom controlled-$Z$ gate, a midpoint atomic qubit memory preserves the entanglement established by whichever photon arrives first while the other QPU continues attempting. If the system coherence time allows a sufficiently long waiting window, the same-attempt requirement is removed and the entanglement rate approaches the linear scaling $Rp$.}
    \label{fig:epair}
  \end{figure}

\begin{figure}[tbp]
    \centering
    \includegraphics[width=\linewidth]{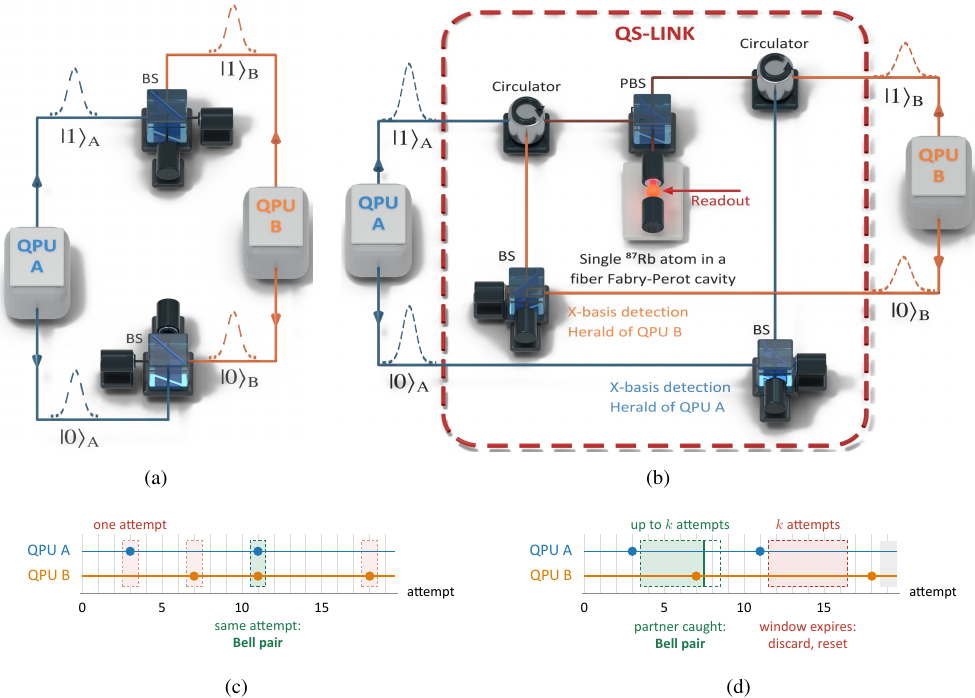}
    \caption{\textbf{Type-II fusion and atomic-qubit-based Bell-state measurements.}
      (a) Linear-optics BSM. Photons from both QPUs must arrive at the interconnect in the same attempt.
      (b) Memory-assisted atom--cavity BSM. For each photon, the $|0_p\rangle$ rail bypasses the cavity, while the $|1_p\rangle$ rail is reflected from it and acquires an atom-state-dependent phase, implementing the controlled-$Z$ gate. A circulator directs the reflected rail to a balanced beam splitter, where it interferes with the bypassed rail. Detection at either output supplies the herald, and the output port identifies the $X$-basis outcome. Atomic read-out completes the BSM after a photon from each QPU has been detected and measured.
    (c) Example detection record for type-II fusion, where both photons must be detected in the same attempt. (d) Example detection record for the memory-assisted protocol. The first herald opens a window of up to $k$ additional attempts in which to obtain the second herald. The second herald closes the window, while expiration triggers a reset.}
    \label{fig:arch}
  \end{figure}

  Quantum teleportation provides a natural route to implementing quantum operations between QPUs over lossy channels: qubits in the two processors are entangled through a repeat-until-success process to create Bell pairs, which are then consumed as needed to transfer quantum states between the QPUs or enact nonlocal gates. Optical photons are natural carriers for distributing such entanglement~\cite{kimble,quic}. They can travel between processors through room-temperature optical fiber without extending the cryogenic or vacuum environment of either processor across the link~\cite{main}. Each processor can entangle a designated communication qubit with an outgoing photonic qubit, which is sent to the interconnect~\cite{simonirvine2003,main}. In the conventional two-photon approach, a type-II fusion gate~\cite{simonirvine2003,browne} performs a Bell-state measurement (BSM) by interfering the photons from the two QPUs on a beam splitter (Fig.~\ref{fig:epair}(a))~\cite{krutyanskiy2023}. A successful two-photon detection pattern heralds the projection of the QPUs' communication qubits onto a well-defined Bell state~\cite{simonirvine2003,main}. Conditioned on both photons arriving, the ideal BSM succeeds with probability 0.5 (Fig.~\ref{fig:arch}(a))~\cite{browne}.

  A major challenge in this field is the typically low per-attempt probability $p$ with which each QPU can generate and deliver a photon to the input of the BSM apparatus. This probability includes the complete source-to-interconnect chain: state preparation of the communication qubit, generation of a photon entangled with this qubit, coupling it into the optical fiber, and propagation loss. With superconducting qubits, the photonic qubit generation also includes microwave-to-optical conversion or the generation of entangled microwave--optical photon pairs (as has been demonstrated by a number of experimental groups~\cite{sahu2022,sahu2023,meesala2024,zhao_transduction,werner2026}, alongside integrated transducers designed for scalable deployment~\cite{weaver2024}). Overall, among cavity-free atomic interfaces, reported end-to-end values of $p$ reach the few-percent level for neutral atoms~\cite{safari2026} and the percent level for trapped ions~\cite{saha,oreilly}, while demonstrated values for superconducting-qubit interfaces remain at or below the percent level, with substantial improvements projected~\cite{mirhosseini,weaver2025}. This low probability presents a tremendous challenge to the linear-optics-based approach for remote entanglement. Beyond the fact that the interference mechanism of the type-II fusion gate requires the photons from the two QPUs to be indistinguishable in all their degrees of freedom (including their temporal envelope), they both need to successfully arrive simultaneously at the beam splitter in the same attempt~\cite{browne}. Together with its probabilistic nature, this means that the success probability of this operation depends quadratically on $p$ and is bounded, even in the ideal case, by $0.5p^2$ per attempt~\cite{browne,calsamiglia2001}. By comparison, the success probability of single-click protocols based on single-rail encoding~\cite{cabrillo1999,hermans2023} scales linearly with $p$, since only one photon is required to herald entanglement. However, suppressing errors from undetected double-excitation events requires very low excitation probabilities, which limits the achievable entanglement rate. These protocols also require stabilization of the relative optical phase accumulated along the long paths from the pump-laser source to the QPUs and from the QPUs to the interconnect~\cite{cabrillo1999,hermans2023}.
  Alternatively, extreme-photon-loss (EPL) distillation can remove these double-excitation errors and cancel this phase while retaining linear scaling with channel efficiency~\cite{campbell2008,dirnegger2026}. However, it requires an additional memory qubit and local two-qubit gates at each QPU, as well as indistinguishable photons and phase stability throughout the waiting time between heralding the first and second pairs~\cite{hermans2023,campbell2008,dirnegger2026}.

  Here, we instead use dual-rail encoding and a quantum memory qubit at the midpoint between the two QPUs to perform the BSM asynchronously~\cite{bhaskar,lei}. This removes the requirement that both photons arrive in the same attempt [Figs.~\ref{fig:epair} and~\ref{fig:arch}(b,d)]. Crucially, the first QPU--memory link must be heralded, because the waiting window begins only once that link is known to have been established.

  In this study, we propose and analyze an inter-QPU entangling protocol in which a single trapped neutral atom implements this memory-assisted BSM through two sequential applications of a robust, symmetric photon--atom controlled-$Z$ (CZ) gate based on the Duan--Kimble cavity-reflection mechanism~\cite{duankimble,reiserer,nagib2024,arwas2026}. Detecting and measuring the first photon heralds successful entanglement between the memory qubit and the communication qubit of one QPU. This entanglement is then preserved while the other QPU continues attempting to send photonic qubits. The protocol succeeds if a photon from the second QPU arrives before the stored atom--QPU link loses coherence. We explicitly account for the additional decoherence caused when a photon interacts with the memory qubit but is not subsequently detected. Finally, measuring the atomic qubit projects the three-qubit Greenberger--Horne--Zeilinger (GHZ) state shared by the two QPUs and the atom onto a well-defined Bell state shared by the two QPUs.

  In the following, we describe the cavity quantum electrodynamics (QED) operations that enable this protocol. We derive the fidelity-limited waiting window for the stored link and the corresponding expected entanglement rate, accounting for coherence decay of the stored link and corruption by unheralded photon interactions. For the atom--cavity implementation considered here, we use the projected parameters of QS-LINK, an integrated atom--cavity system developed by Quantum Source~\cite{arwas2026}. Our model, based on analytical calculations and supported by Monte Carlo simulations, predicts that this system has the potential to enhance the entanglement rate by orders of magnitude over the linear-optics approach using type-II fusion.

  \section{Memory-Assisted Atom--Cavity Interconnect}

  \subsection{Cavity-QED Implementation}
  \label{sec:interface}

  The implementation considered here uses a single $^{87}$Rb atom trapped in a high-finesse fiber Fabry--Perot microcavity resonant with the $D_1$ transition at 795\,nm. The atom serves as the midpoint memory [Fig.~\ref{fig:arch}(b)], and the BSM consists of two reflection-based photon--atom CZ gates followed by optical read-out of the atomic qubit.
  For modular QPU scaling, the main near-term application of quantum interconnects, expected link lengths are tens of meters. At these distances, fiber propagation loss at 795\,nm (which can be as low as 2\,dB/km) is negligible compared with the other inefficiencies that limit $p$. If telecom wavelengths are nonetheless required, photons can be converted locally to 795\,nm at the interconnect with efficiency $\eta>0.5$~\cite{vanleent}. In that case, the conversion efficiency is included in $p$ in the following analysis.

  In our scheme, the QPUs send photons to the interconnect through single-mode optical fibers. For this transmission, they may use time-bin encoding, which is particularly well suited to these links because both bins propagate through the same spatial mode. Phase fluctuations of the long transmission path are slow compared with both the photon pulse duration and the separation between the time bins, and are therefore common to both bins~\cite{beukers2024}.
  This way, the optical link does not require interferometric phase stabilization, only polarization stabilization, which can be performed at a very slow (minutes) timescale\cite{vanleent}.
  At the interconnect, the time-bin qubit can be mapped coherently onto the two spatial paths using a fast optical switch (e.g. a Pockels cell) and a delay matched to the time-bin separation. After this conversion, only the relative phase between the short local paths must be stabilized.

\begin{figure}[tb]
    \centering
    \includegraphics[width=\linewidth]{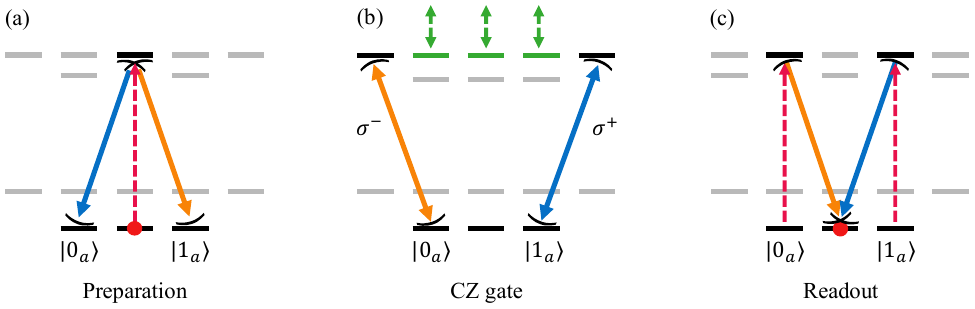}
    \caption{\textbf{Atomic transitions used by the cavity-QED memory.}
    (a) \emph{Preparation.} A $\pi$-polarized pulse (dashed red) excites the atom from the $|F=1,m_F=0\rangle$ ground state. Subsequent decay through the cavity-enhanced $\sigma^-$ and $\sigma^+$ channels (orange and blue, respectively) prepares the atomic qubit in a superposition of $|0_a\rangle$ and $|1_a\rangle$. (b) \emph{CZ gate.} The two atomic qubit states couple to separate excited states through the cavity-enhanced $\sigma^-$ and $\sigma^+$ transitions. A resonant 1324\,nm level-engineering field (green) dresses the central $|F'=2,m_F=0\rangle$ excited state, suppressing unwanted transitions through it and enabling the atomic-state-dependent cavity response required for the gate. (c) \emph{Read-out.} A $\pi$-polarized control field (dashed red) maps the atomic qubit onto an emitted photon while returning the atom to the $|F=1,m_F=0\rangle$ ground state. Adapted from Ref.~\cite{arwas2026}.}
    \label{fig:levels}
  \end{figure}

  We use a symmetric CZ gate based on the Duan--Kimble cavity-reflection gate~\cite{duankimble,reiserer}. The gate acts on a path-encoded dual-rail photonic qubit, with $|0_p\rangle$ and $|1_p\rangle$ represented by a photon in one of two spatial paths. The $|0_p\rangle$ rail bypasses the cavity, whereas a photon in the $|1_p\rangle$ rail is reflected from the cavity and acquires an atom-state-dependent phase, thereby implementing the CZ gate [Fig.~\ref{fig:levels}]. Photons from the two QPUs have orthogonal linear polarizations, allowing their cavity-interacting rails to be combined at a polarizing beam splitter without an active switch between the QPU inputs [Fig.~\ref{fig:arch}(b)]. In the symmetric implementation~\cite{nagib2024,arwas2026}, the gate suppresses errors caused by state-dependent photon loss by coupling both atomic qubit states symmetrically to the cavity. The gate also converts failures such as atom loss or preparation outside the qubit space into photon loss by routing the photon to another output port, thereby eliminating these sources of infidelity~\cite{nagib2024,arwas2026}.  An auxiliary 1324\,nm control field dresses and splits the $|F'=2,m_F=0\rangle$ excited level, suppressing unwanted transitions through it. For the projected QS-LINK parameters, the CZ gate duration is approximately 20\,ns and the photon-survival efficiency is approximately 98.4\%~\cite{arwas2026}.

  Atomic qubit read-out uses vacuum-stimulated Raman adiabatic passage (vSTIRAP) driven by a $\pi$-polarized control pulse~\cite{hennrich,Nisbet-Jones_2011,Mucke_PhysRevA.87.063805}. This pulse maps the atomic qubit onto an emitted photon while returning the atom to the $|F=1,m_F=0\rangle$ ground state. Detecting this photon supplies the final measurement outcome needed to identify the Bell state shared by the communication qubits of the two QPUs [Fig.~\ref{fig:levels}]. The fidelity of atomic state preparation, the two photon--atom CZ gates, and atomic read-out is projected to be 0.98 for successfully heralded events. These operation errors are separate from the memory-induced infidelity budget used below.

  \subsection{Atom-Mediated Bell-State Measurement Protocol}

\begin{figure}[tb]
    \centering
    \includegraphics[width=\linewidth]{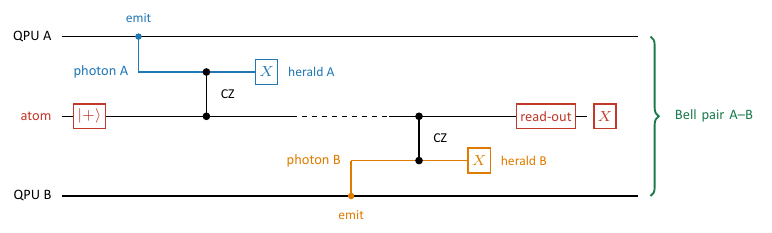}
    \caption{\textbf{Circuit representation of the atom-mediated Bell-state measurement.}
    The two photon--atom CZ gates and photonic $X$-basis measurements establish the two heralds. The dashed portion of the atomic line denotes storage between them, and atomic read-out completes the BSM.}
    \label{fig:protocol}
  \end{figure}

  The interconnect uses the photon--atom CZ gate and atomic read-out described above to perform the memory-assisted BSM between the photonic qubits sent from the two QPUs [Fig.~\ref{fig:protocol}].
  Before the first herald, both QPUs send photons toward the interconnect in each attempt. Each QPU prepares its communication qubit and outgoing photon in the Bell state $(|0_{A,B}\rangle|0_p\rangle+|1_{A,B}\rangle|1_p\rangle)/\sqrt{2}$. Within each attempt, photons from the two QPUs are timed to arrive at the atom in consecutive, nonoverlapping time slots (assumed in the following analysis to be approximately 50\,ns apart).
  Without loss of generality, we denote by QPU-A the QPU whose photon produces the first herald, and by QPU-B the other.

  We denote the photonic $X$-basis measurement outcomes by $m_A$ and $m_B$, with $m=0$ for the $|+_p\rangle$ outcome and $m=1$ for the $|-_p\rangle$ outcome. The protocol consists of the following steps:

  \begin{enumerate}
    \item The atom is prepared in $|+_a\rangle=(|0_a\rangle+|1_a\rangle)/\sqrt{2}$. Detecting the first photon in the $X$ basis after the photon--atom CZ gate heralds the state $(|0_A\rangle|+_a\rangle+(-1)^{m_A}|1_A\rangle|-_a\rangle)/\sqrt{2}$ shared by the communication qubit of QPU-A and the atom. If neither photon is detected in an attempt, the atom is re-prepared in the $|+_a\rangle$ state before the next attempt.

    \item The second herald can occur in the same attempt as the first. However, if only one herald is obtained in that attempt, QPU-A preserves its communication qubit while only QPU-B continues emitting, making up to $k$ additional attempts using the same qubit--photon Bell state. Detecting a photon from QPU-B in the $X$ basis after the CZ gate heralds a three-qubit GHZ state shared by the atom and the communication qubits of both QPUs.

    \item {\sloppy Finally, the atomic qubit is measured in the $X$ basis by mapping its state onto a cavity photon via vSTIRAP and measuring the photon's polarization in the horizontal/vertical ($H/V$) linear-polarization basis. The $|+_a\rangle$ read-out outcome projects the QPU qubits onto $(|0_A\rangle|0_B\rangle+(-1)^{m_A+m_B}|1_A\rangle|1_B\rangle)/\sqrt{2}$, whereas the $|-_a\rangle$ outcome gives $(|0_A\rangle|1_B\rangle+(-1)^{m_A+m_B}|1_A\rangle|0_B\rangle)/\sqrt{2}$, up to a global phase. The three measurement outcomes therefore identify the Bell state shared by the two QPUs.\par}
  \end{enumerate}

  If no herald signal from QPU-B is obtained within the $k$ attempts of the waiting window, the measurement results are discarded and the protocol restarts.
  In Sec.~\ref{sec:window} we derive the maximal allowed number of attempts $k$ from the average memory-induced infidelity budget.
  Because the photons from the two QPUs interact with the atom sequentially and never interfere, they need not be indistinguishable. They need only be centered near the $F=1\rightarrow F'=2$ transition of the $D_1$ line and lie within the cavity-enhanced bandwidth of the atom.
  Their temporal envelopes may otherwise differ~\cite{aqua,arwas2026}. For the parameters considered here, pulse durations as short as $10\,\mathrm{ns}$ satisfy this condition~\cite{arwas2026}.

  \section{Rate and Fidelity Model}
  \label{sec:model}

  In each attempt, repeated at rate $R$, each QPU delivers a photonic qubit to the interconnect input with probability $p$. Conditional on reaching the input, the photon interacts with the atom with probability $\etain$, giving the photon--atom interaction probability $\ptilde=\etain p$. Following the interaction, the photon is detected with probability $\etaout$. For convenience, we define the per-QPU heralding
  probability as $q=\ptilde\etaout$. We retain $\etain$ and $\etaout$ as separate parameters
  because losses before and after the photon--atom interaction have
  qualitatively different consequences. Loss before the interaction only
  reduces the entanglement-generation probability. By contrast, loss after
  the interaction, in step~2, leaks which-state information about the stored
  atom--QPU-A entangled state to the environment and therefore induces
  dephasing of that state. Together, $\etain$ and $\etaout$ account for the finite CZ gate efficiency, optical losses, and detector efficiency. In the following we assume identical parameters and statistically independent photon-delivery events for the two QPUs.

  \subsection{Entanglement Rate with a Finite Memory Window}
  At the start of the protocol, both QPUs attempt to supply photonic qubits
  to the interconnect. If no photon is detected in an attempt, the state of both QPUs and the atom is reset. The probability of a same-attempt full success, namely that
  both photons successfully reach the interconnect and generate heralding
  events in the same attempt, is \(q^2\). For
  \(q \ll 1\), obtaining exactly one herald is more likely and occurs with probability $2q(1-q)$. This event establishes entanglement between the atom and the corresponding QPU qubit, and opens a window of up to $k$ additional attempts in which to obtain the second herald from the other QPU. The probability of obtaining the second herald before the window closes is $w=1-(1-q)^k$. As shown in Appendix~\ref{app:rate}, the resulting entanglement rate is
  \begin{equation}
    \Gamma=\etaro R\,
    \frac{q^2+2q(1-q)w}{1+2(1-q)w}.
    \label{eq:rate}
  \end{equation}
  Here $\etaro$ is the atomic read-out efficiency. The numerator gives the probability that a full protocol cycle produces the two photon detections required to entangle the QPUs. The denominator gives the mean number of attempts used in that cycle, including any waiting for the second detection. If the atomic readout fails, all results are discarded and the protocol restarts from the beginning.

  When the probability of obtaining the second herald within the waiting window
  is small, \(qk \ll 1\), we have \(w \simeq qk\) and therefore
  $\Gamma \simeq \eta_{\rm ro}(2k+1)q^2R$.
  In this limit the rate remains quadratic in \(q\), but is enhanced by a factor of
  \(2k+1\). This factor comprises the same-attempt contribution together with
  two sets of \(k\) additional attempts, corresponding to the two possible
  choices of the QPU that generates the first herald. In the opposite limit of an effectively unbounded waiting window,
  \(w \to 1\), and for \(q \ll 1\), we obtain
  $\Gamma \simeq \eta_{\rm ro}\frac{2}{3}qR$.
  The factor \(2/3\) follows from the mean time required to obtain both
  heralds. Before the first herald, both QPUs are active, giving a combined
  emission-attempt rate of \(2R\) and a mean waiting time of approximately
  \(1/(2qR)\). After the first herald, only QPU-B remains active, yielding an
  additional mean waiting time of \(1/(qR)\). The total mean waiting time is
  therefore \(3/(2qR)\), which gives the factor \(2/3\) in the entanglement
  rate.

  \subsection{Fidelity-Limited Memory Window}
  \label{sec:window}

  While awaiting the second herald, the entanglement between the atom and QPU-A is exposed to two sources of error. The first is the decoherence of the stored link, which we approximate by an exponential with effective coherence time $\Tc$, giving a factor $e^{-1/(R\Tc)}$ per attempt. For Gaussian atomic decay and exponential QPU decay, with $T_{2,\mathrm{atom}}^*\geq T_{2,\mathrm{QPU}}^*$, the effective time is chosen to approximate their product over the accepted waiting window. Appendix~\ref{app:qpu-coherence} discusses this approximation, while Appendix~\ref{app:validation} validates the resulting entanglement-rate predictions using Monte Carlo simulations. The second source of error arises when a photon from QPU~B interacts with the atom but is subsequently lost before detection. This unheralded interaction occurs with probability $\ptilde(1-\etaout)$ per attempt and completely dephases the atom, thereby destroying the stored entanglement.

  When calculating coherence decay, we neglect the approximately 50\,ns offset between photon arrivals within an attempt. This offset is much shorter than both the atomic and QPU coherence times used in Sec.~\ref{sec:scaling}.
  Within this exponential approximation, if the second herald occurs $j$ attempts after the first, the expected memory-induced fidelity of the completed pair is
  \begin{equation}
    F_j=\frac{1}{2}\left[1+
    \left(\frac{1-\ptilde}{1-q}e^{-1/(R\Tc)}\right)^j\right].
    \label{eq:pair-fidelity}
  \end{equation}
  The ratio $(1-\ptilde)/(1-q)$ is the probability that no photon interacted with the atom, conditioned on the absence of a herald, and therefore accounts for unheralded photon interactions. The exponential describes the decay of the stored link's coherence. Averaging over all immediate and delayed completions accepted within a window $k$ gives
  \begin{equation}
    \bar F(k)=
    \frac{1+2\sum_{j=1}^{k}(1-q)^jF_j}
    {1+2\sum_{j=1}^{k}(1-q)^j}.
    \label{eq:average-fidelity}
  \end{equation}
  We denote the allowed average memory-induced infidelity by $\delta$ and choose $k$ as the largest integer for which $1-\bar F(k)\leq\delta$. This criterion determines the waiting windows used for the main-text figures and quoted rates.

  Combining the rate and fidelity models reveals a simple scaling. For $q\ll1$, $w\simeq1-e^{-qk}$, so Eq.~(\ref{eq:rate}) shows that the normalized rate $\Gamma/(qR)$ depends on the waiting window through $qk$. We consider a waiting window that contains many attempts but remains short compared with the mean waiting time for a second herald. Expanding Eqs.~(\ref{eq:pair-fidelity}) and~(\ref{eq:average-fidelity}) to leading order in the memory-induced infidelity then gives (Appendix~\ref{app:fidelity})
  \begin{equation}
    qk \simeq
    \frac{4\delta\etaout}
    {1-\etaout+1/\Nph},
    \qquad
    \Nph=\ptilde R\Tc.
    \label{eq:product}
  \end{equation}
  Here $\Nph$ is the mean number of photons supplied by a given QPU that interact with the atom during one effective link coherence time. In this limit, $\Nph$ captures the dependence of the normalized rate $\Gamma/(qR)$ on $\ptilde$, $R$, and $\Tc$, while $qR$ sets the absolute entanglement rate.

  The coherence-limited short-window regime admits a simple scaling law. In this regime, Eq.~(\ref{eq:product}) gives $qk\simeq4\delta\etaout\Nph$, and substitution into the short-window rate yields, apart from steps associated with integer changes in $k$,
  \begin{equation}
    \Gamma\simeq
    \underbrace{\etaro q^2R}_{\text{same-attempt}}
    +\underbrace{\frac{8\delta\etaro\etaout^2}{\Tc}\Nph^2}_{\text{memory-assisted}}.
    \label{eq:quadratic-rate}
  \end{equation}
  For interfaces where $\ptilde R\ll1/\Tc$, increasing $\ptilde R$ then provides a double benefit: it raises the incoming herald rate and increases the probability that a partner photon interacts before the stored link decoheres. Consequently, the memory-assisted contribution grows quadratically with $\ptilde R$, whereas increasing $\Tc$ produces linear growth.

  \section{Results}
  \label{sec:results}

  \subsection{Entanglement-Rate Scaling and Enhancement}
  \label{sec:scaling}

\begin{figure}[tb]
    \centering
    \includegraphics[width=\linewidth]{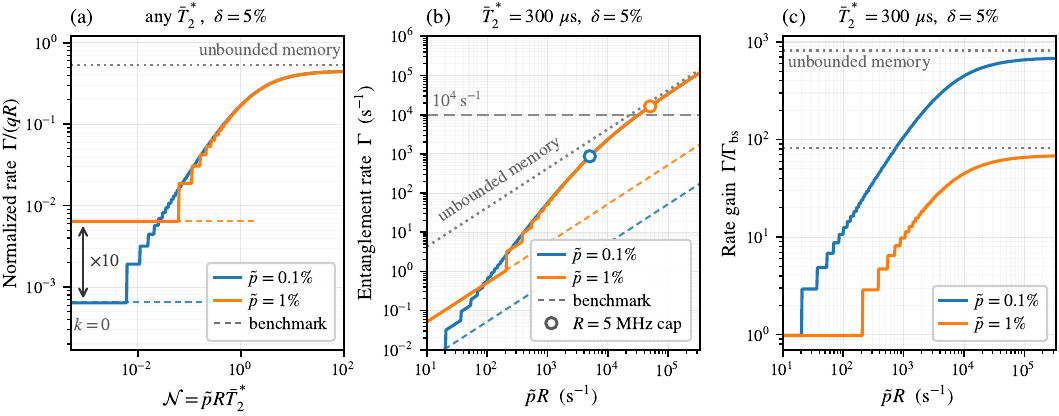}
    \caption{\textbf{Scaling, absolute rate, and gain.}
    The waiting window is selected for an average memory-induced infidelity budget of $\delta=5\%$, with $\etaout=\etaro=0.8$. (a) Normalized entanglement rate versus $\Nph=\ptilde R\Tc$ at $\ptilde=0.1\%$ and $1\%$. (b) Absolute rate versus $\ptilde R$ at $\Tc=300\,\mu\mathrm{s}$. Open circles mark $R=5$\,MHz and the horizontal dashed line marks $10^4$\,s$^{-1}$. (c) Gain over type-II fusion. Dashed curves show the benchmark, gray dotted curves the unbounded-window limit, and steps arise from integer-valued $k$. The plotted variable satisfies $\ptilde=\etain p$, with the projected memory-input efficiency $\etain=0.9$. The benchmark assumes indistinguishable photons and $\eta=\eta_{\mathrm{det}}=0.92$.}
    \label{fig:unified}
  \end{figure}

  Figure~\ref{fig:unified} summarizes how the entanglement rate of the memory-assisted interconnect depends on $\Nph$ and $\ptilde R$, together with its gain over type-II fusion. Panel (a) shows the entanglement rate normalized by the per-QPU herald rate, $\Gamma/(qR)$. At small $\Nph$, the infidelity budget permits no additional attempt, so $k=0$ and only same-attempt heralds contribute. As additional attempts become admissible, the memory-assisted contribution grows and the curves for different $\ptilde$ gradually converge. At large $\Nph$, unheralded photon interactions limit the waiting-window length, and the normalized rate approaches a plateau below the unbounded-window limit.

  For comparison, the entanglement rate for type-II fusion is
  \begin{equation}
    \Gamma_{\mathrm{bs}}=\tfrac{1}{2}R\eta^2p^2,
    \label{eq:bs}
  \end{equation}
  where $\eta$ is the probability that a photon present at the interconnect input is detected. For a best-case reference, we assume indistinguishable photons and take $\eta=\eta_{\mathrm{det}}=0.92$, where $\eta_{\mathrm{det}}$ is the detector efficiency.

  Atomic coherence times of up to $2\,\mathrm{ms}$ can be achieved with a small guiding magnetic field~\cite{langenfeld2020,seubert2026efficient}. We therefore conservatively consider $T_{2,\mathrm{atom}}^*=1\,\mathrm{ms}$ and $T_{2,\mathrm{QPU}}^*=300\,\mu\mathrm{s}$, the latter consistent with demonstrated superconducting-qubit coherence times~\cite{wang2022transmon}. Over short storage times, Gaussian atomic decay contributes only a small correction to the QPU coherence, motivating the choice $\Tc=300\,\mu\mathrm{s}$ for panels (b) and (c). For QPU platforms with Gaussian coherence decay, the same characteristic coherence times give a slower initial decay of the stored link. The exponential benchmark is then conservative over the short waiting windows considered here, allowing higher entanglement rates at the same infidelity budget and otherwise identical protocol parameters. These panels show the entanglement rate and rate gain as functions of $\ptilde R$, the rate at which photons from each QPU interact with the atom. At low rates of interaction, where the waiting window is limited by the finite coherence time, the memory-assisted rate in panel (b) grows quadratically with $\ptilde R$, as predicted by Eq.~(\ref{eq:quadratic-rate}). At higher rates of interaction, unheralded photon interactions limit the waiting window, and the growth becomes linear.

  The solid curves differ at low rates because $k$ can take only integer values, but converge at higher rates, and the memory-assisted rate then depends mainly on $\ptilde R$. The type-II fusion benchmark instead scales as $Rp^2$. Since $\ptilde=\etain p$, at fixed $\ptilde R$ and $\etain$ the benchmark remains proportional to $\ptilde$. Reducing $\ptilde$ by a factor of ten while increasing $R$ by the same factor therefore leaves the memory-assisted rate nearly unchanged in this regime but reduces the benchmark by a factor of ten. This is why panel (c) shows the larger gain for the lower value of $\ptilde$.

  The comparison can also be expressed in terms of the attempt rate required to reach a chosen entanglement rate. Although this requirement depends on the fault-tolerant architecture, recent studies place entanglement rates of order $10^4\,\mathrm{s}^{-1}$ in a useful regime across several platforms~\cite{menssen2026,sinclair2025,knollmann2026,pattison2025,marqversen2026,weaver2025}. At $\ptilde=1\%$, reaching this rate with the parameters of Fig.~\ref{fig:unified} requires $R_{\mathrm{mem}}\simeq3.2$\,MHz for the memory-assisted interconnect and $R_{\mathrm{bs}}\simeq190$\,MHz for type-II fusion. The memory therefore reduces the required attempt rate by nearly a factor of sixty. Recent ion-based interfaces have demonstrated percent-level photon probabilities~\cite{saha,oreilly}, including operation at a megahertz-scale repetition rate~\cite{oreilly}, placing the memory-assisted requirement within reach of realistic interface capabilities. Type-II fusion would instead require operation in the hundreds-of-megahertz regime.

  \subsection{Rate--Fidelity Trade-off and Coherence Limit}

\begin{figure}[tb]
    \centering
    \includegraphics[width=\linewidth]{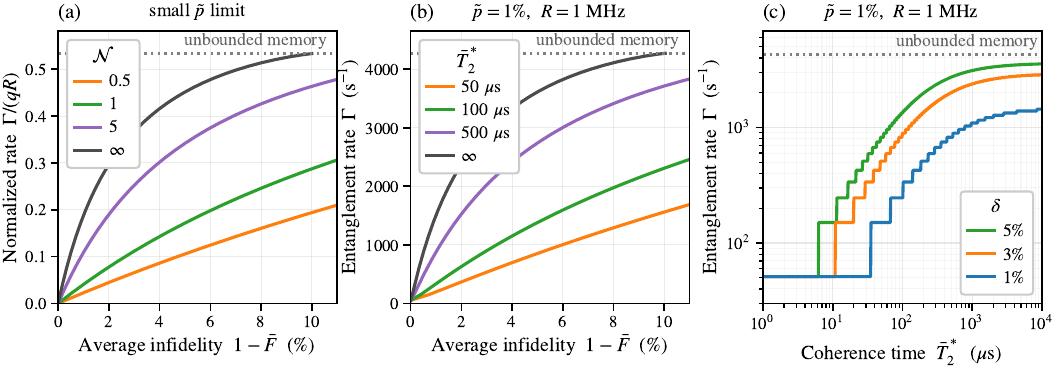}
    \caption{\textbf{Rate--fidelity trade-off and effect of link coherence.}
    Parameters are $\etaout=\etaro=0.8$. (a) Normalized rate versus average memory-induced infidelity in the small-$\ptilde$ limit for several $\Nph$. (b) Absolute trade-off at $\ptilde=1\%$ and $R=1$\,MHz for several coherence times. Corresponding colors in (a) and (b) have the same $\Nph$. (c) Rate versus $\Tc$ at the same $\ptilde$ and $R$ for $\delta=1\%$, 3\%, and 5\%. Increasing the integer window length $k$ moves the operating point along the curves in (a) and (b) and through the steps in (c). Dotted lines show the unbounded-window rate.}
    \label{fig:tradeoff}
  \end{figure}

  Once the first herald has been obtained, the waiting-window length determines how many further attempts can be made to obtain the second. A longer window increases the probability of completing the Bell pair, and hence the entanglement rate, at the cost of greater average memory-induced infidelity. Figure~\ref{fig:tradeoff}(a) shows this trade-off in dimensionless form. At a fixed average infidelity, a larger $\Nph$ gives a higher normalized rate because the second herald is more likely to occur within the link coherence time.

  Panel (b) shows the same trade-off in absolute units for $\ptilde=1\%$ and $R=1$\,MHz. For these parameters, $\Nph=\Tc/(100\,\mu\mathrm{s})$, so corresponding colors in panels (a) and (b) represent the same number of photon--atom interactions per coherence time.

  Panel (c) fixes the allowed average memory-induced infidelity $\delta$ and shows how the entanglement rate changes with $\Tc$. At short coherence times, increasing $\Tc$ allows more attempts to obtain the second herald and raises the entanglement rate. The improvement slows once unheralded photon interactions become the main limitation. Equation~(\ref{eq:product}) predicts that coherence decay and unheralded photon interactions contribute equally at $\Nph=1/(1-\etaout)=5$ for $\etaout=0.8$. For the parameters of panel (c), this corresponds to $\Tc=500\,\mu\mathrm{s}$. Beyond this point, increasing $\Tc$ reduces the coherence-decay contribution but not the error caused by unheralded photon interactions.

  With $\etaout=0.8$, an unbounded waiting window would produce an average infidelity of $10\%$ from unheralded photon interactions alone. The $1\%$, $3\%$, and $5\%$ budgets shown in panel (c) are all below this value. The selected waiting window therefore remains finite even for long coherence times, causing the rates to plateau below the unbounded-window limit.

  \section{Discussion and Outlook}
  \label{sec:discussion}

  The central practical result is that a midpoint memory is most useful when photons are scarce. When the infidelity budget permits $k\geq1$, the entanglement created by a herald from one QPU can be retained when the other QPU fails in the same attempt. Two quantities capture this behavior: $qR$ sets how quickly heralds are obtained from each QPU, while $\Nph=\ptilde R\Tc$ governs the fraction of first heralds that lead to a completed Bell pair within the memory window. For interfaces with $\ptilde R\ll1/\Tc$, increasing $\ptilde R$ improves both quantities, so the memory-assisted contribution grows as $(\ptilde R)^2$, whereas type-II fusion at fixed $p$ grows only linearly with $R$. Increasing $\Tc$, by contrast, improves only the fraction completed. In this regime, the memory provides the strongest leverage for improvements in photon delivery or repetition rate.

  Once $\Nph$ is large enough that unheralded photon interactions dominate the memory error, further increases in $\Tc$ yield diminishing returns. Improving transmission and detection after the atom is then especially valuable: it raises the herald rate while reducing the probability that an interaction corrupts the stored link without producing a herald. This loss channel also explains why a stricter infidelity budget can keep the optimal window finite even for a very long-lived memory. Atomic read-out efficiency has a simpler role, multiplying the entanglement rate without changing the selected window.

  Existing interfaces occupy markedly different parts of this parameter space. Trapped-ion systems have demonstrated $1\,\mu\mathrm{s}$ attempt cycles with percent-level single-photon success probabilities~\cite{oreilly}, directly reaching the $1\%$, 1\,MHz regime considered here. Links to superconducting QPUs require microwave-to-optical transduction, and recent work has analyzed their performance using existing and near-term transducers without assuming deterministic conversion~\cite{weaver2025}. For these links, conversion and optical-coupling losses reduce $p$, the transducer bandwidth and duty cycle limit the achievable attempt rate $R$, and transducer-added noise contributes additional infidelity~\cite{sahu2022,weaver2024,zhao_transduction}. Appendix~\ref{app:added-noise} extends the model to include this noise and illustrates how increasing the conversion probability can come at the expense of a shorter useful memory window.

  The fidelity model isolates the errors accumulated while the first link is stored. It takes the qubit--photon states supplied by the two QPUs as inputs and does not describe how they are generated. Errors in preparing these input states, together with QPU-side errors beyond the dephasing already included in the effective link coherence time, must therefore be included separately in an end-to-end fidelity budget. For the atom--cavity implementation, we assume reasonable trapping lifetimes on the order of seconds, far longer than the protocol timescales considered here, and neglect atom loss and reloading. Downtime required for motional ground-state cooling is not included in the entanglement-rate model.
  Although the waiting window is selected using an ensemble-average fidelity, the detection record retains pair-specific information: the delay between the two photon detections provides an estimate of the memory-induced error for that pair. This information can be passed to soft-information decoders~\cite{li2024soft,pattison2021soft}.

  Finally, this protocol is particularly suitable for interconnecting QPUs
  based on different physical platforms: because entanglement is mediated
  asynchronously by the atom and does not rely on two-photon interference,
  photons from the two QPUs need not be mutually indistinguishable; they need
  only be resonant with the \(795\,\mathrm{nm}\) atomic transition and fall
  within the bandwidth of the cavity-enhanced atom--photon interface. The same interface could therefore support heterogeneous quantum networks linking trapped-ion and neutral-atom QPUs with superconducting QPUs equipped with microwave-to-optical transducers, requiring only wavelength conversion, which can be done at reasonable efficiencies\cite{vanleent}. Using time-bin encoding for transmission and converting it locally to path-encoded dual rail confines interferometric phase stabilization to the short optical paths within the interconnect rather than the full QPU--interconnect links. Demonstrating the complete protocol in such a heterogeneous quantum network would be an important next step toward a practical modular architecture for quantum computation.


  \appendix
  \renewcommand{\theHsection}{app.\Alph{section}}

  \section{Derivation of the Entanglement Rate}
  \label{app:rate}

  To derive Eq.~(\ref{eq:rate}), we calculate the probability of obtaining both heralds in one cycle and the average number of attempts that cycle uses. A cycle consists of the initial attempt and, if exactly one herald is obtained, the subsequent waiting window.

  Both heralds are obtained either in the initial attempt, with probability $q^2$, or after exactly one initial herald followed by a second within the window, with probability $2q(1-q)w$. Their combined probability is therefore
  \[
    q^2+2q(1-q)w,
  \]
  where $w=1-(1-q)^k$ is the probability of obtaining the second herald within the $k$ additional attempts.

  To calculate the average cycle duration, we first count the attempts made within a waiting window. The first additional attempt is always made; the second is made only if the first fails, and so on. The $j$th additional attempt is therefore made with probability $(1-q)^{j-1}$. Summing these probabilities gives the average number of additional attempts:
  \begin{equation}
    \sum_{j=1}^{k}(1-q)^{j-1}
    =\frac{1-(1-q)^k}{q}
    =\frac{w}{q}.
    \label{eq:windowlength}
  \end{equation}

  Every cycle uses one initial attempt, while a waiting window opens with probability $2q(1-q)$. The average number of attempts per cycle is thus
  \[
    1+2q(1-q)\frac{w}{q}=1+2(1-q)w.
  \]

  At attempt rate $R$, the average cycle duration is this number divided by $R$. Multiplying the probability of obtaining both heralds by the read-out efficiency $\etaro$, then dividing by the average cycle duration, gives
  \[
    \Gamma=\etaro R\,
    \frac{q^2+2q(1-q)w}{1+2(1-q)w},
  \]
  which is Eq.~(\ref{eq:rate}).

  \subsection{Useful Limits}

  The exact result provides three useful checks. When $k=0$, $w=0$ and $\Gamma=\etaro Rq^2$, so only pairs for which both heralds occur in the same attempt contribute to the entanglement rate. For $qk\ll1$, $w\simeq kq$ and
  \begin{equation}
    \Gamma\simeq\etaro(2k+1)q^2R.
    \label{eq:shortwindow}
  \end{equation}
  The factor $2k+1$ includes the same-attempt contribution and two sets of $k$ additional attempts, one for each choice of which QPU produces the first herald. In the unbounded-window limit, $w\to1$ and
  \begin{equation}
    \Gamma_\infty
    =\etaro R\,\frac{q(2-q)}{3-2q}
    \xrightarrow{q\ll1}
    \etaro\frac{2}{3}qR.
    \label{eq:longwindow}
  \end{equation}
  For $q\ll1$, the first herald arrives after approximately $1/(2q)$ attempts and the second arrives after approximately $1/q$ additional attempts. The resulting mean cycle length, $3/(2q)$ attempts, gives the factor $2/3$.

  \section{Average Bell Pair Fidelity}
  \label{app:fidelity}

  This appendix derives the fidelity relations in Eqs.~(\ref{eq:pair-fidelity}) and~(\ref{eq:average-fidelity}) and their leading-order reduction in Eq.~(\ref{eq:product}). The rate includes every pair completed within the selected window, so the infidelity constraint is imposed on the probability-weighted average over those pairs.

  If the pair is completed on the $j$th additional attempt, the absence of a second herald in the initial attempt and in the following $j-1$ attempts gives $j$ opportunities for an unheralded photon interaction. Conditional on the absence of a herald, the probability that no photon interacted with the atom is $(1-\ptilde)/(1-q)$. Combining the survival probability over these $j$ events with coherence decay during the waiting time gives Eq.~(\ref{eq:pair-fidelity}).

  Obtaining both heralds in the same attempt has probability $q^2$ and fidelity $F_0=1$. A pair completed on waiting attempt $j$ has probability $2q^2(1-q)^j$ and fidelity $F_j$, where the factor of two accounts for either QPU producing the first herald. Dividing the fidelity-weighted sum of these events by their total probability gives Eq.~(\ref{eq:average-fidelity}). The read-out efficiency cancels from this conditional average because a failed read-out causes the attempt to be discarded rather than reducing the fidelity of an accepted pair.

  The selected window is the largest integer $k$ satisfying $1-\bar F(k)\leq\delta$. For the parameter range considered here, the left-hand side increases monotonically with $k$, so the exact threshold can be found directly and inserted into Eq.~(\ref{eq:rate}). A requirement on every accepted pair would instead impose $1-F_k\leq\delta$ on the longest delay and would generally select a smaller window. All reported rates use the average criterion.

  To obtain Eq.~(\ref{eq:product}), define the leading-order coherence loss per attempt
  \[
    \lambda=-\ln\left(\frac{1-\ptilde}{1-q}\right)
    +\frac{1}{R\Tc}
    \simeq (1-\etaout)\ptilde+\frac{1}{R\Tc},
  \]
  where the approximation holds for $q,\ptilde\ll1$. When $qk\ll1$, delayed completions are approximately uniform across the window. Including the contribution from pairs completed in the same attempt gives
  \[
    1-\bar F(k)\simeq
    \frac{\lambda}{2}\frac{k(k+1)}{2k+1}.
  \]
  In the many-attempt limit, $1\ll k\ll1/q$, setting this expression to $\delta$ gives
  \[
    \delta\simeq\frac{k}{4}
    \left[(1-\etaout)\ptilde+\frac{1}{R\Tc}\right].
  \]
  Using $q=\etaout\ptilde$ and $\Nph=\ptilde R\Tc$ then yields
  \[
    qk\simeq
    \frac{4\delta\etaout}
    {1-\etaout+1/\Nph},
  \]
  which is Eq.~(\ref{eq:product}). This continuous approximation exposes the scaling. Every numerical operating point uses the exact integer criterion based on Eq.~(\ref{eq:average-fidelity}).

  \section{Effect of Transducer-Added Noise}
  \label{app:added-noise}

  Microwave-to-optical transducers can add noise photons to the optical output~\cite{sahu2022,weaver2024,zhao_transduction}. We denote the number of added photons per signal photon in the relevant optical mode by $N_{\mathrm{add}}$ and assume that signal and noise photons experience the same losses and cannot be distinguished by the atom or detector. To obtain a simple model of this effect, we work in the low-occupancy limit, where a given channel in each attempt contains either vacuum, one signal photon, or one noise photon. We neglect simultaneous signal-and-noise photons and multiple-noise-photon events, which enter at higher order in the corresponding per-attempt occupancies. The total photon--atom interaction and herald probabilities are then
  \begin{equation}
    \tilde p_{\mathrm{tot}}
    =\ptilde(1+N_{\mathrm{add}}),
    \qquad
    q_{\mathrm{tot}}
    =\etaout\tilde p_{\mathrm{tot}}
    =q(1+N_{\mathrm{add}}).
    \label{eq:noise-probabilities}
  \end{equation}

  Conditional on a herald, the probability that the detected photon was a signal photon is
  \begin{equation}
    \Pr(\mathrm{signal}\mid\mathrm{herald})
    =\frac{1}{1+N_{\mathrm{add}}}.
  \end{equation}
  A completed pair requires two heralds, so the probability that both arose from signal photons is $1/(1+N_{\mathrm{add}})^2$. We model a noise-generated herald as removing the coherence between the two components of the target Bell state while preserving their populations. A pair with at least one such herald then has fidelity $1/2$. Added noise therefore introduces the infidelity floor
  \begin{equation}
    \delta_{\mathrm{add}}
    =\frac{1}{2}\left[
      1-\frac{1}{(1+N_{\mathrm{add}})^2}
    \right].
    \label{eq:noise-floor}
  \end{equation}

  To include this contribution in the memory model, $\ptilde$ and $q$ are replaced by $\tilde p_{\mathrm{tot}}$ and $q_{\mathrm{tot}}$ in the rate and fidelity expressions. Noise photons that interact with the atom but are not detected are therefore included among the unheralded interactions that dephase the stored link. If $\delta_{\mathrm{mem}}$ denotes the resulting memory-induced infidelity, the total infidelity is
  \begin{equation}
    \delta_{\mathrm{tot}}
    =\delta_{\mathrm{add}}
    +(1-2\delta_{\mathrm{add}})\delta_{\mathrm{mem}}.
  \end{equation}
  Here, $\delta$ bounds the combined infidelity from memory errors and transducer-added noise. The effective budget available for memory-induced errors is therefore
  \begin{equation}
    \delta_{\mathrm{eff}}
    =\frac{\delta-\delta_{\mathrm{add}}}
    {1-2\delta_{\mathrm{add}}}.
    \label{eq:effective-budget}
  \end{equation}
  The pump power cannot be increased beyond the point at which added noise alone exhausts the infidelity budget. We denote this value by $P_{\max}$, for which $\delta_{\mathrm{add}}=\delta$. For $\delta=5\%$, this occurs at $N_{\mathrm{add}}\simeq5.4\%$.

\begin{figure}[tb]
    \centering
    \includegraphics[width=0.86\linewidth]{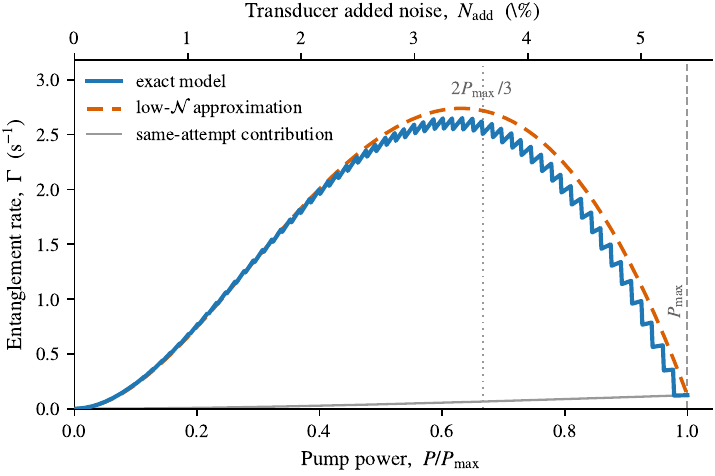}
    \caption{\textbf{Effect of transducer-added noise.}
    Entanglement rate versus normalized pump power, with the corresponding added noise shown on the upper axis. The solid curve uses the exact integer waiting-window criterion, the dashed curve shows the low-$\Nph$ approximation, and the gray curve is the same-attempt contribution. The fidelity budget sets the maximum allowed pump power $P_{\max}$. The marker at $2P_{\max}/3$ shows the limiting estimate for the optimum. Steps in the exact curve arise from integer changes in $k$. The conversion curve is normalized to $p=0.1\%$ at $C=1$, with $C(P_{\max})=0.18$. Other parameters are $\delta=5\%$, $R=1\,\mathrm{MHz}$, $\Tc=300\,\mu\mathrm{s}$, $\etain=0.9$, and $\etaout=\etaro=0.8$.}
    \label{fig:added-noise}
  \end{figure}

  Increasing the transducer pump power increases the conversion probability but can also introduce additional noise~\cite{sahu2022}. To illustrate this trade-off, we take the cooperativity and added noise to increase linearly with pump power and use the standard cavity electro-optic conversion curve~\cite{xu2021}:
  \begin{equation}
    C\propto P,
    \qquad
    p\propto\frac{4C}{(1+C)^2},
    \qquad
    N_{\mathrm{add}}\propto P.
    \label{eq:pump-model}
  \end{equation}
  In the low-cooperativity regime, $C\ll1$, these relations give $p\propto P$. In the coherence-limited, short-window regime, the entanglement rate is approximately
  \begin{equation}
    \Gamma\simeq
    \etaro q_{\mathrm{tot}}^2R
    \left(1+8\delta_{\mathrm{eff}}R\Tc\right).
    \label{eq:noise-rate-approx}
  \end{equation}
  Over the allowed pump range, we approximate the remaining memory-error budget as
  \begin{equation}
    \delta_{\mathrm{eff}}\simeq
    \delta\left(1-\frac{P}{P_{\max}}\right).
    \label{eq:pump-budget}
  \end{equation}
  When the memory-assisted term dominates, $p\propto P$ therefore gives
  \begin{equation}
    \Gamma\propto P^2\left(1-\frac{P}{P_{\max}}\right),
  \end{equation}
  which gives the limiting estimate $P=2P_{\max}/3$ for the optimal pump power. Figure~\ref{fig:added-noise} compares the rate approximation in Eq.~(\ref{eq:noise-rate-approx}) with the exact integer-window calculation.

  The optimal operating point is therefore not determined by conversion probability alone. Increasing the pump power produces more signal photons, but the accompanying noise leaves a smaller budget for memory-induced errors and shortens the useful memory window. The conversion probability and added noise must consequently be evaluated at the same operating point when determining the attainable entanglement rate.

  \section{Numerical Consistency Check}
  \label{app:validation}

  We validate the analytical model using Monte Carlo simulations of the protocol. The analytical curves use the waiting window $k$ selected with the exponential fidelity expression in Eq.~(\ref{eq:pair-fidelity}). For the Monte Carlo simulations, $k$ is instead selected with the exact fidelity expression in Eq.~(\ref{eq:pair-fidelity-full}), which includes Gaussian atomic decay and exponential QPU decay. Both choices use the average fidelity in Eq.~(\ref{eq:average-fidelity}) and the same infidelity budget $\delta$.

  The simulation then records the number of completed pairs and the total number of attempts for the selected waiting window, giving
  \[
    \Gamma_{\mathrm{MC}}=\etaro R\,\frac{N_{\mathrm{success}}}{N_{\mathrm{attempts}}}.
  \]
  Each plotted point uses $10^5$ completed pairs.

\begin{figure}[tb]
    \centering
    \includegraphics[width=\linewidth]{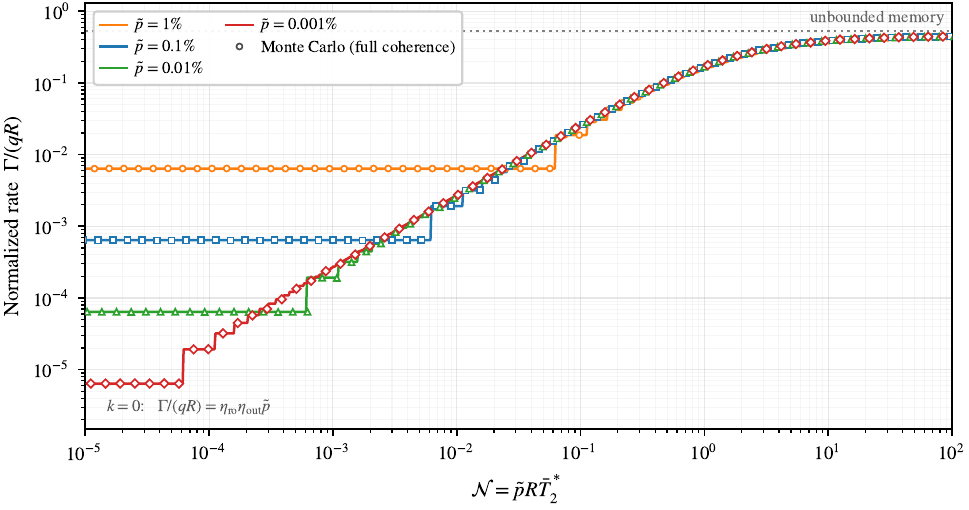}
    \caption{\textbf{Monte Carlo comparison with the exponential model.}
    Normalized rate versus $\Nph=\ptilde R\Tc$. Curves use the exponential model, while symbols show Monte Carlo results using the full Gaussian--exponential coherence. Each selects its waiting window at the same average-infidelity budget. Parameters are $T_{2,\mathrm{atom}}^*=1\,\mathrm{ms}$, $T_{2,\mathrm{QPU}}^*=\Tc=300\,\mu\mathrm{s}$, $\etaout=\etaro=0.8$, and $\delta=5\%$. Error bars denote one standard error. The dotted curve is the unbounded-window limit.}
    \label{fig:mcrate}
  \end{figure}

  Figure~\ref{fig:mcrate} compares the simulated rates with the analytical predictions over three decades of $\ptilde$ and seven decades of $\Nph$. At $\ptilde=1\%$ and $R=1\,\mathrm{MHz}$, the exact fidelity expression selects $k=41$, giving $\Gamma=2.32\times10^3\,\mathrm{s}^{-1}$, compared with $k=42$ and $\Gamma=2.35\times10^3\,\mathrm{s}^{-1}$ for the exponential approximation, a difference of $1.3\%$.

  \begin{samepage}
    \section{Gaussian Atomic Coherence and the Exponential Approximation}
    \label{app:qpu-coherence}

    Atomic coherence times of up to $1\,\mathrm{ms}$ can be achieved with a small guiding magnetic field~\cite{langenfeld2020}. For independent Gaussian atomic decay and exponential QPU decay, the stored-link coherence is
    \begin{equation}
      C_{\mathrm{link}}(t)=\exp\!\left[-\left(\frac{t}{T_a}\right)^2-\frac{t}{T_q}\right],
      \label{eq:full-coherence}
    \end{equation}
    where $T_a=T_{2,\mathrm{atom}}^*$ and $T_q=T_{2,\mathrm{QPU}}^*$ are the respective $1/e$ coherence times.
  \end{samepage}

  \begin{samepage}
    Replacing the exponential coherence factor in Eq.~(\ref{eq:pair-fidelity}) by $C_{\mathrm{link}}(j/R)$ gives the memory-induced fidelity
    \begin{equation}
      F_j^{\mathrm{full}}=\frac{1}{2}\left[1+
        \left(\frac{1-\ptilde}{1-q}\right)^j
      \exp\!\left(-\frac{j}{RT_q}-\left(\frac{j}{RT_a}\right)^2\right)\right].
      \label{eq:pair-fidelity-full}
    \end{equation}
    Substituting this expression into Eq.~(\ref{eq:average-fidelity}) gives the corresponding average fidelity.
  \end{samepage}

\begin{figure}[tb]
    \centering
    \includegraphics[width=\linewidth]{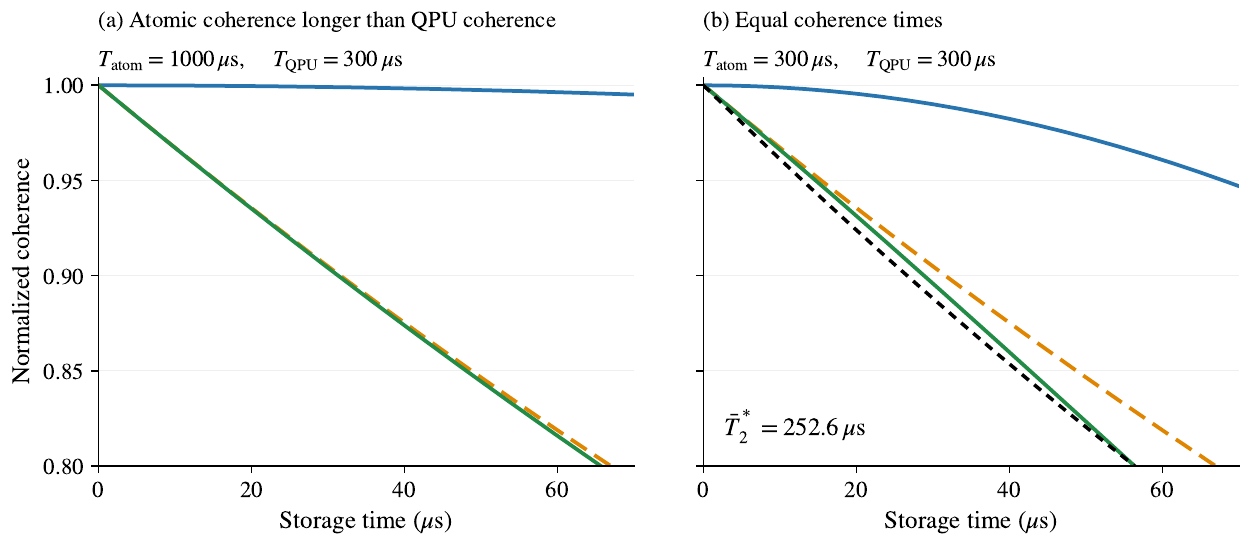}
    \caption{\textbf{Gaussian atomic coherence and the exponential approximation.}
    Atomic coherence (blue solid), QPU coherence (orange dashed), and their product (green solid) for (a) $T_a=1\,\mathrm{ms}$ and $T_q=300\,\mu\mathrm{s}$, and (b) $T_a=T_q=300\,\mu\mathrm{s}$. The additional black dashed line in (b) shows a conservative exponential approximation with $\Tc=252.6\,\mu\mathrm{s}$.}
    \label{fig:coherence-approx}
  \end{figure}

  For $T_a=1\,\mathrm{ms}$ and $T_q=300\,\mu\mathrm{s}$, Gaussian atomic decay has little effect over short storage times, so the product closely follows the QPU exponential [Fig.~\ref{fig:coherence-approx}(a)]. When the two coherence times are equal, a slightly faster exponential remains close to the product while giving a conservative approximation over the displayed range [Fig.~\ref{fig:coherence-approx}(b)].

  The figure shows coherences between 0.8 and 1. For pure dephasing, $F=(1+C)/2$. In the many-attempt, short-window limit, a $5\%$ average-infidelity budget permits roughly $10\%$ infidelity for the latest accepted pairs, corresponding to coherence near 0.8.

  \subsection*{Disclosures}
  J.R., N.K., G.A., Y.A., Y.J., T.K., and B.D. are employees of Quantum Source Labs Ltd.

  \subsection*{Code, Data, and Materials Availability}
  The code and simulation data supporting the results presented in this paper are available from the authors upon reasonable request.

  \subsection*{Acknowledgments}
  We thank Nimrod Shenker for the helpful discussions. We acknowledge support from the Israel Innovation Authority. B.D. acknowledges support from the Israel Science Foundation, the U.S.--Israel Binational Science Foundation, the Minerva Foundation, and the United States Army Research Office (Grant No.~W911NF-24-1-0392). B.D. holds the Dan Lebas and Roth Sonnewend Professorial Chair of Physics.


\end{spacing}
\end{document}